\documentclass[%
 aps,
 amsmath,amssymb,
 reprint,%
]{revtex4-1}

\usepackage{graphicx}
\usepackage{dcolumn}
\usepackage{bm}
\usepackage[left]{lineno}

\usepackage{lipsum}
\usepackage[utf8]{inputenc}
\usepackage[T1]{fontenc}
\usepackage{mathptmx}
\usepackage{etoolbox}
\usepackage{xcolor}
\makeatletter
\def\@email#1#2{%
 \endgroup
 \patchcmd{\titleblock@produce}
  {\frontmatter@RRAPformat}
  {\frontmatter@RRAPformat{\produce@RRAP{*#1\href{mailto:#2}{#2}}}\frontmatter@RRAPformat}
  {}{}
}%

\makeatother

\begin{document}

\preprint{AIP/123-QED}
%
%
%
%
%

\title{Giant optical oscillator strengths in perturbed hexagonal germanium}

\author{Abderrezak Belabbes$^{1}$}
\email {abderrezak.belabbes@uni-jena.de}
\author{Friedhelm Bechstedt $^{2}$}
\author{Silvana Botti$^{2}$}
\affiliation{$^1$
\mbox{Department of Physics, Sultan Qaboos University, P.O. Box 36, PC 123, Muscat, Oman}}
\affiliation{$^2$ 
\mbox{Institut f\"ur Festk\"orpertheorie and -optik, Friedrich-Schiller-Universit\"at Jena, Max-Wien-Platz 1, 07743 Jena, Germany}}


\begin{abstract}
We present \emph{ab initio} calculations of electronic and optical properties of perturbed hexagonal germanium and demonstrate that it is a superior material for active
optoelectronic devices in the infrared spectral region. It is known that perfect lonsdaleite Ge is a
pseudodirect semiconductor, i.e., with direct fundamental band gap but almost vanishing oscillator strength for the lowest-energy optical transitions. Perturbing the system by replacing a Ge atom in the unit cell with a Si atom boosts of the oscillator strength at the minimum direct gap by orders of magnitude, with a concurrent blue shift of the interband distances.
This effect is mainly due to the increased $s$ character of the lowest conduction band because of the  perturbation-induced wave function mixing. A purely structural modification of the lonsdaleite unit cell of hexagonal Ge yields as well increased optical oscillator strengths, 
but their magnitude significantly depends on the actual details of the atomic geometry. In particular, moderate tensile uniaxial strain can induce an inversion of the order of the two lowest conduction bands, immediately leading to an extremely efficient enhancement of optical transitions.
In general, chemical and/or structural perturbations of the lonsdaleite lattice are shown to be the key to make hexagonal germanium suitable for light emitting devices.
\end{abstract}

\maketitle

\section{\label{sec1}Introduction}

Diamond silicon (Si) is the key material of semiconductor industry. However, since it is an indirect band-gap semiconductor, it cannot emit light efficiently. Therefore, Si cannot be employed in active photonic devices or in optical intrachip communication. Modifications of this fundamental semiconductor are desirable to increase its light emission and absorption efficiency and make it suitable for optoelectronic applications \cite{Ball:2001,Atabaki.Moazeni.ea:2018:N}. Mixing Si with germanium (Ge) keeps the compatibility of the resulting
Si$_x$Ge$_{1-x}$ alloys with CMOS processing , but this modification still leads to indirect semiconductors. Only alloying in combination with a variation the crystal structure from cubic diamond (3C) to hexagonal lonsdaleite (2H)
\cite{Raffy.Furthmueller.ea:2002:PRB} allows for efficient light emission. This has been recently demonstrated for nanostructured hexagonal (hex) Ge-rich alloys \cite{Fadaly:2020:N}. \

First-principles electronic structure calculations for bulk lonsdaleite Ge \cite{Roedl.Furthmueller.ea:2019:PRM} find that this material has a direct band gap at the Brillouin zone (BZ) center $\Gamma$ but with an extremely
small oscillator strength associated to the optical transition from the valence band maximum (VBM) into the lowest conduction band
minimum (CBM). As a consequence, 2H-Ge can be classified as a pseudodirect semiconductor. Replacing a fraction of the Si atoms with Ge to obtain the alloy hex-Si$_x$Ge$_{1-x}$, with the 2H bond pattern, 
induces a strong decrease by three orders of magnitudes of radiative lifetime until reaching about 50\% Si content \cite{Borlido-PRB-2021}. 
The direct-indirect transition of the band gap character as a function of Si molar fraction $x$ has been recently addressed by several studies  \cite{Borlido-PRB-2021,Cartoixa-nanolett-2017,Bao-Nanyun-JAP-2021,Wang-Zhen-APL-2021}. The optical properties of these alloys have been calculated from first principles, including quasiparticle and excitonic effects, for Si-rich hexagonal alloys within the virtual crystal approximation \cite{Cartoixa-nanolett-2017} and for 2H-Ge \cite{Bao-Nanyun-JAP-2021}. However, the quantum mechanical reasons of the forbidden optical transitions at the direct band gap of 2H-Ge have remained unexplored. An explanation of how alloying in Ge-rich systems or perturbations of the atomic arrangements influence the onset transitions is still missing. As a result, there is a need to investigate how modifications of 2H-Ge can transform it from a pseudodirect semiconductor to a direct band-gap material.    

The drastic changes in the optoelectronic properties going from pure 2H-Ge to Ge-rich hex-Si$_x$Ge$_{1-x}$ alloys ask for
a deeper microscopic understanding. To this end, in this Letter we calculate variations of the electronic band structure
and compare optical matrix elements for dipole transitions with photon energies $\hbar\omega$ near the absorption edge for hexagonal alloys with about 25\% silicon and pure 2H-Ge. More precisely, we consider the hexagonal four-atom cell and its BZ and investigate, besides
pure lonsdaleite germanium, also the crystal obtained by replacing one Ge atom with a Si one. The latter atomic arrangement describes a hexagonal Si$_{0.25}$Ge$_{0.75}$ alloy with an almost homogeneous distribution of Si substitutions over the entire crystal. To distinguish between chemical and structural perturbations of the starting 2H-Ge crystal, we perform calculations also for atomic arrangements of only Ge atoms but with the equilibrium geometry of hexagonal Si$_{0.25}$Ge$_{0.75}$. Finally, as an example of stronger structural
perturbation, we consider the geometry obtained by applying 1.8\% tensile uniaxial strain along the $c$-axis on 2H-Ge and the resulting changes of the electronic structure.

\section{\label{sec3} Results and Discussions}

We optimized the considered hex-Ge and hex-SiGe cells and calculated the corresponding electronic and optical properties using density functional theory (DFT). The computational details are discussed in section~\ref{sec2}.

The incorporation of 25\% of Si induces a shrinking of the lattice in comparison with 2H-Ge, yielding for Si$_{0.25}$Ge$_{0.75}$ lattice constants of $a=3.946$ (3.996)~{\AA} and
$c=6.155$ (6.592)~{\AA} with ratio $c/a=1.651$ (1.650) $>\sqrt{8/3}$ and an internal-cell parameter $u=0.374$ (0.374) $\stackrel{<}{\sim}3/8$
(the values in parenthesis are those for 2H-Ge~\cite{Roedl.Furthmueller.ea:2019:PRM}). The space group symmetry is lowered from P6$_3$/mmc ($C^4_{6v}$) to P3m1 ($C^1_{3v}$), accompanied by a point group change from 6/mmm ($D_{6h}$) to $\bar{6}$m2 ($D_{3h}$). The $u$ parameter for hex-Si$_{0.25}$Ge$_{0.75}$  is an average of the values $u=0.370/0.378$ along the vertical Si-Ge/Ge-Ge bond in the unit cell. Symmetry-conserving uniaxial strain leads to lattice parameters $a=3.987$~{\AA}, $c=6.712$~{\AA}, and $u=0.372$. The elongation of the unit cell along the
$c$-axis is accompanied by a shrinking of the cell in the perpendicular directions.

DFT calculations of Kohn-Sham band structures are known to drastically underestimate the interband transition energies and band gaps when semi-local exchange correlation (XC) functionals
are employed \cite{Bechstedt:2015:Book}. In view of that, to simulate the quasiparticle (QP) corrections to the DFT band structure
due to electron or hole excitations, we apply an improved XC functional by Tran and Blaha \cite{Tran.Blaha.ea:2007:JoPCM,Tran.Blaha:2009:PRL}: the modified Becke-Johnson (MBJ) potential\cite{Becke.Johnson:2006:TJoCP} with the correlation term in the local density approximation (LDA), called MBJLDA functional. This functional has been
proved to lead to accurate QP band energies for semiconductors, and in particular for Si and Ge crystals
\cite{Roedl.Furthmueller.ea:2019:PRM,Borlido-PRB-2021,Laubscher.Kuefner.ea:2015:JoPCM,Borlido-Aull:JCTC-2019,Borlido-Jonathan-npj2020}. 

For the purpose of illustrating the precision of the MBJLDA electronic energies, the Bloch bands at $\Gamma$ are also computed using another approximate QP method, namely the HSE06 hybrid functional of Heyd, Scuseria and Ernzerhof
(HSE) \cite{Heyd.Scuseria.ea:2003,Heyd.Scuseria.ea:2006:JoCP}. In Table~\ref{tab1} we can observe a rather good agreement
between the energies calculated within the two approximate QP methods for both hex-Si$_{0.25}$Ge$_{0.75}$ and 2H-Ge. The
differences of valence-band energies are smaller than 10~meV. Even in the case of unoccupied bands
the maximum deviations for the alloy remain below 0.1~eV. The optical dipole-matrix elements are computed
in the longitudinal gauge \cite{Gajdos.Hummer.ea:2006:PRB} and presented in terms of Bloch matrix elements
$P^\alpha_{cv}({\bf k})=\langle c{\bf k}|p_\alpha|v{\bf k}\rangle$ of the $\alpha th$ Cartesian component $p_\alpha$
of the momentum operator for vertical transitions between valence band states $|v{\bf k}\rangle$ and conduction
band states $|c{\bf k}\rangle$ with energies $\varepsilon_\nu({\bf k})$ $(\nu=c,v)$. Excitonic effects are neglected since they are known to be weak at the absorption onset of germanium, independently of the Ge polymorph, because of the strong dielectric screening \cite{Bao-Nanyun-JAP-2021,Malone-PRB-2010}.

\begin{figure}[h]
\includegraphics[width=0.9\columnwidth]{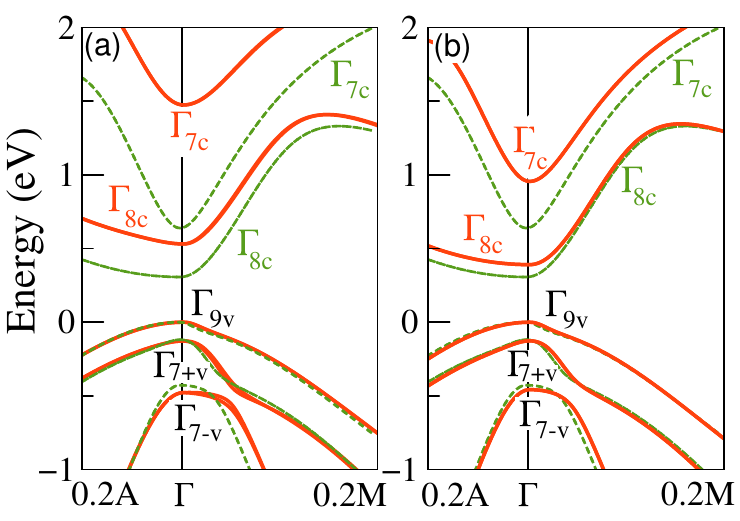} 
\caption{Band structures around the $\Gamma$ point (red solid lines) of (a) hexagonal Si$_{0.25}$Ge$_{0.75}$ and (b) structurally perturbed Ge -- arranged in the atomic geometry
of optimized hexagonal Si$_1$Ge$_3$ -- versus ${\bf k}$-points along high-symmetry lines in the hexagonal BZ. The bands are compared with the bands (green dashed lines) of pure
2H-Ge. The uppermost $\Gamma_{9v}$ valence band is used as energy zero}. The band classification of the lonsdaleite symmetry group is also applied for the alloyed and structurally perturbed hex-Ge. 
\label{fig1}
\end{figure}

The QP band structures of hex-Si$_{0.25}$Ge$_{0.75}$ and of pure 2H-Ge with the lattice constants and the atomic positions of hex-Si$_{0.25}$Ge$_{0.75}$ are displayed in the panels (a) and (b), respectively, of Figs.~\ref{fig1}. The electronic bands are plotted as a function of the vector ${\bf k}$ along high-symmetry lines in the BZ, considering only a small region of the BZ around the $\Gamma$ point and zooming on the energy interval close to the direct band gap. We apply for clarity the symmetry classification of the Bloch states derived for lonsdaleite also to hex-Si$_{0.25}$Ge$_{0.75}$ and the distorted 2H-Ge cells, despite their symmetry reduction. The corresponding energy levels
and band dispersions, i.e., the effective masses, near $\Gamma$ are given in Table~\ref{tab1}. In the case of 2H-Ge the energies at $\Gamma$ perfectly agree with previous calculations using the same approach \cite{Roedl.Furthmueller.ea:2019:PRM}. This statement holds for the characteristic parameters of the splittings of the $p$-like valence
bands, the crystal-field splitting $\Delta_{cf}=0.27$~eV and the spin-orbit splittings $\Delta^{\|}_{so}=0.28$~eV and
$\Delta^\bot_{so}=0.27$~eV. The effective masses are however slightly increased, apart from those
of $\Gamma_{7\pm v}$ along $\Gamma$M. This is due to the strong sensitivity of the effective
masses to the size of the ${\bf k}$-space volume around the $\Gamma$-point considered to perform the parabolic fit of the band
dispersion. The bands in Fig.~\ref{fig1}(b), calculated for pure Ge using the atomic positions of the hex-SiGe cell, lead to only small variations with respect to the band structure of 
 2H-Ge. This especially holds for the valence bands. The conduction bands are slightly shifted toward higher energies as a consequence of the structural perturbation of the lonsdaleite lattice. Considering the reduction of the lattice constants, our findings are in qualitative agreement with results for a hydrostatic volume decrease shown in Ref.~\cite{Suckert.Roedl.ea:2021:PRM}.   \

\begin{figure}[h]
\includegraphics[width=6cm]{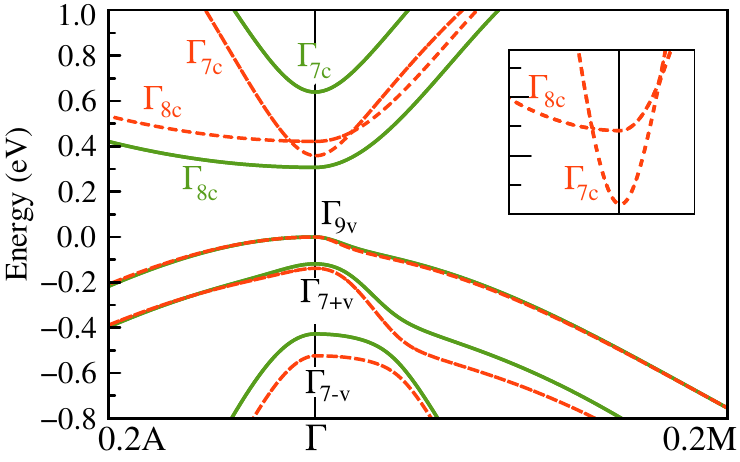} 
\caption{Band structure of uniaxially strained 2H-Ge (red dashed lines) in comparison with the unstrained crystal (green solid lines). The inset illustrates the band inversion at $\Gamma$. }
\label{fig2}
\end{figure}

 The largest shift happens for the second conduction band. More precisely, the
changes of the atomic positions induce band shifts of 0.311 ($\Gamma_{7c}$), 0.061 ($\Gamma_{8c}$), 0.000
($\Gamma_{9v}$), -0.010 ($\Gamma_{7+v}$), and -0.032 ($\Gamma_{7-v}$)~eV with respect to 2H-Ge
(see Table~\ref{tab1}). Totally different bands are instead obtained for uniaxially strained hex-Ge (1.8\% tensile strain). In particular,
this specific cell deformation leads to an inversion of the $\Gamma_{7c}$ and $\Gamma_{8c}$ conduction bands as visible in Fig.~\ref{fig2}. This result is 
in agreement with earlier predictions~\cite{Suckert.Roedl.ea:2021:PRM}. The resulting band energies are 0.358 ($\Gamma_{7c}$),
0.421 ($\Gamma_{8c}$), 0.000 ($\Gamma_{9v}$), -0.138 ($\Gamma_{7+v}$), and -0.523~eV ($\Gamma_{7-v}$).
Because of the pure $s$-character of the $\Gamma_{7c}$ state, its energy drastically moves down toward the VBM, while the $p$-containing $\Gamma_{8c}$ conduction band only slightly shifts toward higher energies.
This inversion of the two lowest conduction bands is in agreement with results of other strain studies for 2H-Ge \cite{Inaoka.Furukawa.ea:2015:JoAP,Mellaerts.Afanasaev.ea:2021:AAMI}
 and also with the indirect-direct transition in 3C-Ge under tensile uniaxial strain \cite{Suckert.Roedl.ea:2021:PRM}. 

\begin{table*}
\caption{Modification of band distances and band dispersion near $\Gamma$ in hex-Si$_{0.25}$Ge$_{0.75}$ with respect to
2H-Ge (values given in parenthesis). The state labels of pure 2H-Ge are applied. 
The valence-band maximum is used
as energy zero. The band energies computed with the HSE06 functional are listed in addition for the purpose of comparison. Three high-symmetry directions between the BZ center and boundary perpendicular ($\Gamma$M,$\Gamma$K) and parallel ($\Gamma$A) to the $c$-axis are chosen to represent effective band masses. All effective masses are given in units of the free-electron mass $m$.}
	\begin{ruledtabular}
\begin{tabular}{cccccc}
 & \multicolumn{2}{c}{position (eV)}&\multicolumn{3}{c}{effective mass ($m$)} \\ 
  \cline{2-3} \cline{4-6} \\
   State&MBJLDA&HSE06& $\Gamma \to\ $ M & $\Gamma\to\ $K  & $\Gamma\to\ $A\\ 
 \hline
 $\Gamma_{7c}$    & 1.473 (0.639)      & 1.559 (0.641)      & 0.126 (0.055) & 0.132 (0.059) & 0.068 (0.048) \\
$\Gamma_{8c}$    & 0.530 (0.306)      & 0.561 (0.295)      & 0.094 (0.093) & 0.099 (0.095) & 0.362 (1.098) \\
$\Gamma_{9v}$    & 0.000 (0.000)      & 0.000 (0.000)      & 0.155 (0.092) & 0.311 (0.233) & 0.548 (0.516) \\
$\Gamma_{7+v}$  & -0.127 (-0.118)   & -0.138 (-0.127)   & 0.088 (0.072) & 0.087 (0.064) & 0.250 (0.120) \\
$\Gamma_{7-v}$   & -0.477 (-0.427)   & -0.486 (-0.451)   & 0.495 (0.242) & 0.419 (0.226) & 0.065 (0.048) 
\label{tab1}
\end{tabular}
\end{ruledtabular}
\end{table*}

Mixing Si with Ge, in this case replacing 25\% of the Ge atoms in the hexagonal cell, tends to increase significantly the band
distances (see Fig.~\ref{fig1}(a)), in agreement with the reduction of the atomic size from the covalent radius of 1.2~{\AA} (Ge)
to 1.11~{\AA} (Si) \cite{Sargent-Welch:1980:Book}. The resulting structural modification may be interpreted as a compressive 
internal hydrostatic pressure of
about 2.3\% on the Ge lattice which opens the gap and interband distances. While this effect is small on the valence bands,
the energy distances between conduction bands and their distances to the VBM are significantly
increased, as it can be observed in Fig.~\ref{fig1}(a) and Table~\ref{tab1}. The comparison with the bands of pure 2H-Ge with atoms arranged
as in hex-Si$_{0.25}$Ge$_{0.75}$, shown in Fig.~\ref{fig1}(b), proves that the valence bands are  negligibly affected by this simple structural perturbation. 
However, the conduction-band shifts toward higher energies are 
significantly enhanced by chemical perturbation, i.e., the
replacement of a Ge atom by a Si one. The fundamental gap $E_g=\Gamma_{8c}-\Gamma_{9v}$ is opened from
0.31 to 0.53~eV. The splitting $\Gamma_{7c}-\Gamma_{8c}$ of the two lowest conduction bands dramatically increases
from 0.33 to 0.94~eV. The gap opening is in agreement with
photoluminescence (PL) measurements \cite{Fadaly:2020:N} that indicate a shift of the PL peak position from 0.35~eV for pure Ge to 0.67~eV for Si$_{0.35}$Ge$_{0.65}$ in strong correlation with the fundamental gap variation in
Table~\ref{tab1}. The wavelength $\lambda=3.54$~$\mu$m corresponding to the fundamental gap of 2H-Ge is decreased to
1.85~$\mu$m in hex-Si$_{0.25}$Ge$_{0.75}$. Incorporation of additional Si atoms may further reduce this wavelength toward the communication
wavelength of 1.55~$\mu$m. The corresponding gap $E_g\approx0.8$~eV should be reached for Si$_x$Ge$_{1-x}$
with a composition $x\approx0.56$ applying linear gap interpolation. Similar values have been 
 predicted elsewhere
\cite{Fadaly:2020:N,Borlido-PRB-2021}.

The modifications of band dispersion in Fig.~\ref{fig1}, described around $\Gamma$
by the effective masses in Table~\ref{tab1}, do not show a unique trend with the incorporation of Si into the Ge lattice.
The uppermost conduction band $\Gamma_{7c}$ and the three valence bands $\Gamma_{9v}$, $\Gamma_{7+v}$, and
$\Gamma_{7-v}$ exhibit an increase of the effective masses. Only the lowest conduction band $\Gamma_{8c}$ maintains its dispersion along the direction
perpendicular to the $c$-axis but the effective electron mass parallel to the $c$-axis is reduced. This behavior is
probably related to the increasing repulsion between the two lowest conduction bands with rising $x$, which also explains
the increase of the energy distance $\Gamma_{7c}-\Gamma_{8c}$.

\begin{figure}[h]
\includegraphics[width=7cm]{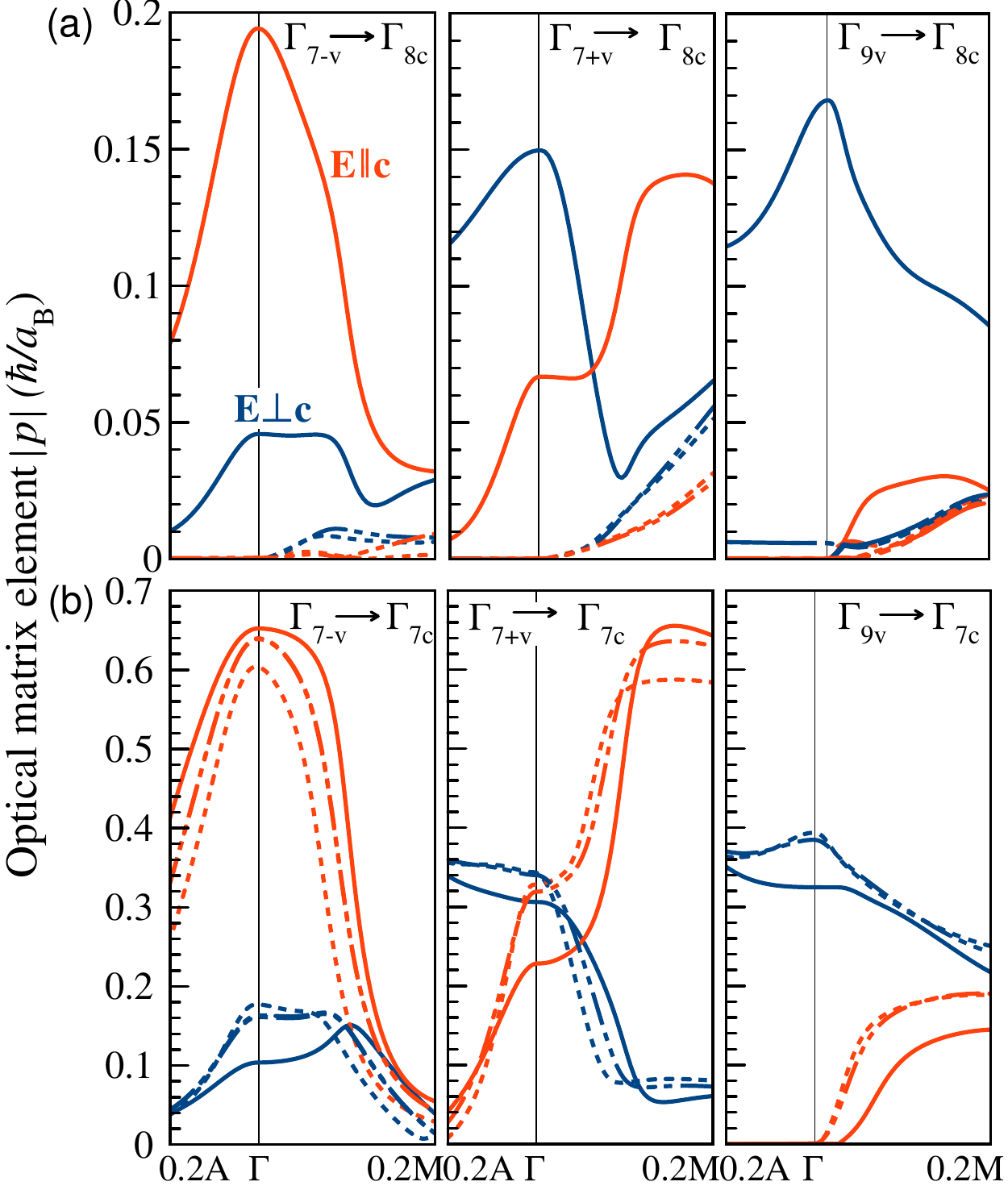}
\caption{Dipole transition matrix elements for optical transitions between the three highest valence bands $\Gamma_{9v}$,
$\Gamma_{7+v}$ and $\Gamma_{7-v}$ and (a) the first lowest conduction band $\Gamma_{8c}$, as well as (b) the second lowest
conduction band $\Gamma_{7c}$ for the hex-Si$_{0.25}$Ge$_{0.75}$ alloy (solid lines), lonsdaleite Ge (dotted lines) and 
the structurally perturbed hexagonal Ge (dot-dashed lines). The curves are blue (red) for light polarization perpendicular (parallel) to
the $c$-axis. }
\label{fig3}
\end{figure}
\begin{figure}[h]
\includegraphics[width=7cm]{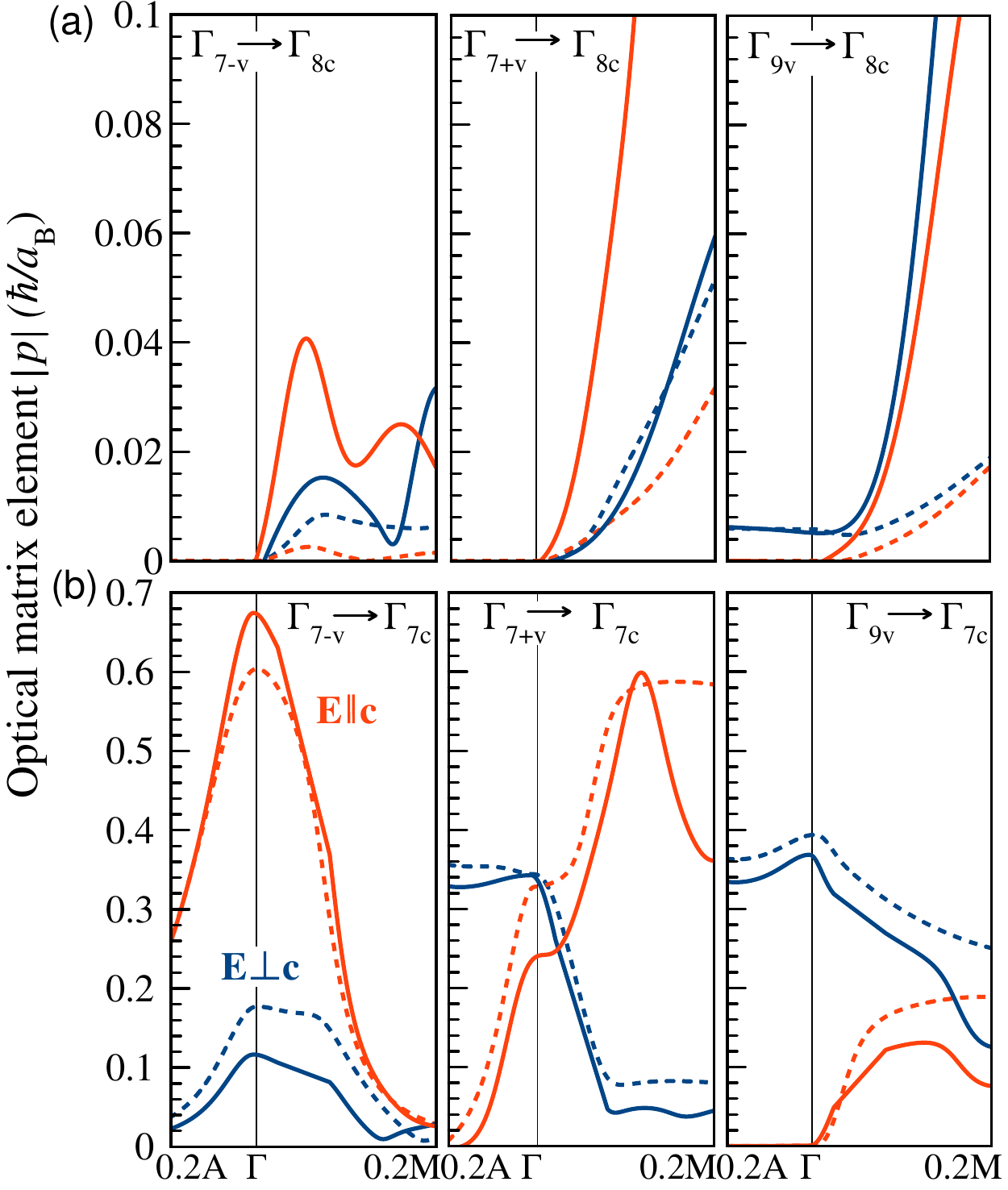} 
\caption{Dipole transition matrix elements as in Fig.~\ref{fig3} but for hex-Ge under a tensile strain of 1.8\% (solid lines),
compared with those of unstrained 2H-Ge (dotted lines). }
\label{fig4}
\end{figure}

Without considering modifications in the band occupation, optical spectra $I^\alpha(\omega)$ such as absorption or luminescence are determined by expressions of the type
\begin{eqnarray}\label{eq1}
I^\alpha(\omega)\sim\frac{1}{V}\sum_{\bf k}
\sum_{c=\Gamma_{7c},\Gamma_{8c}}\sum_{v=\Gamma_{9v},\Gamma_{7+v},\Gamma_{7-v}}
\left|P^\alpha_{cv}({\bf k})\right|^2 \nonumber \\
 \times \delta\left(\hbar\omega-\varepsilon_c({\bf k})+\varepsilon_v({\bf k})\right)
\end{eqnarray}
for light-polarization direction $\alpha$ and photon energy $\hbar\omega$. The optical transition matrix elements
$|P^\alpha_{cv}({\bf k})|$ are plotted in Fig.~\ref{fig3} for all direct transitions between the three highest valence bands
and the two lowest conduction bands along the high-symmetry directions $\Gamma$M and $\Gamma$A (see Fig.~\ref{fig1})
 in pure lonsdaleite Ge, hexagonal Si$_{0.25}$Ge$_{0.75}$ alloy and, for comparison, in
hex-Ge with atomic positions determined by the alloy geometry. In Fig.~\ref{fig4} we present the optical transition matrix elements for hex-Ge under a tensile uniaxial strain of 1.8\% with respect to the ideal lonsdaleite geometry.

In the case of lonsdaleite Ge the curves in Fig.~\ref{fig3} agree with previous results \cite{Roedl.Furthmueller.ea:2019:PRM}. At $\Gamma$ they respect the
selection rules imposed by group theory \cite{Tronc.Kitaev.ea:1999:pssb}. All transitions from the $p$-like valence bands
into the $s$-like second conduction band $\Gamma_{7c}$ in Fig.~\ref{fig3}(b) are allowed, independently of the light
polarization direction. Only for light polarized parallel to the $c$-axis, optical transitions
from the uppermost valence band $\Gamma_{9v}$ are forbidden.
The situation is completely different for optical transitions involving the lowest conduction band $\Gamma_{8c}$, as it can be observed 
in Fig.~\ref{fig3}(a). All transitions at $\Gamma$ are dipole-forbidden. The only exception is the transition
$\Gamma_{9v}\rightarrow\Gamma_{8c}$ for light polarization parallel to the $c$-axis in agreement with group theory
\cite{Tronc.Kitaev.ea:1999:pssb}. However, the optical strength of the latter transition is by orders of magnitude, more precisely a factor of about 70,
smaller than that of the $\Gamma_{9v}\rightarrow\Gamma_{7c}$ transition. The origin of this disappointing result can be
understood within the band-folding picture: the lowest conduction band of diamond Ge at two $L$ points is folded
onto the $\Gamma$ point in the hexagonal BZ. Despite the modification of the atomic stacking in the hexagonal lattice compared with 
 the diamond one, the strong $sp$-character of the $\Gamma_{8c}$ states together with the accompanying vanishing
oscillator strength between $p$-type valence and conduction bands is mainly conserved. The orbital character of the
lowest conduction band is illustrated by orbital and site projections of the $\Gamma_{8c}$ wave function in Table~\ref{tab2}.
Each of the four Ge atoms in the lonsdaleite cell contribute equally and show a 70\%, 20\%, 10)\% orbital character of
$s$. $p_z$, $d_{z^2}$-type, respectively.

Modifying pure 2H-Ge by replacing one Ge atom with Si to obtain hex-Si$_{0.25}$Ge$_{0.75}$, as shown in Fig.~\ref{fig3}(b), leads to only small variations of the optical matrix elements
of the transitions from the valence bands to the second-lowest $\Gamma_{7c}$ conduction band. The situation is totally
different for transitions into the lowest $\Gamma_{8c}$ conduction band in Fig.~\ref{fig3}(a). Apart from the 
$\Gamma_{9v}\rightarrow\Gamma_{8c}$ band transition with light polarization parallel to the $c$-axis, strong optical transitions appear in the alloyed system. Their dipole strengths
are only smaller by factors of 2--3 compared to transitions to the $\Gamma_{7c}$ band. Thereby the averaged $s$ contribution to the conduction
band wave function is slightly increased (see Table~\ref{tab2}). However, the four atoms in the Si$_1$Ge$_3$ unit cell give
rise to drastically different contributions to the total wave function. According to Table~\ref{tab2}, when the two Ge atoms lie above
each other, bonded along the $c$-axis, the $s$-character of the "$\Gamma_{8c}$" wave function considerably increases up to 92\%.
Therefore strong intra-atomic contributions to the oscillator strengths are possible. Small lattice perturbations, such
as due to the substitution of one Ge atom in lonsdaleite by an isoelectronic Si atom together with the accompanying symmetry
reduction of the atomic positions, internal strain and stronger mixing-in of $s$-orbitals into the lowest conduction band induce
drastic changes in the optical oscillator strengths. The lowest optical transition $\Gamma_{9v}\rightarrow\Gamma_{8c}$ becomes
really dipole-allowed for in-plane polarization with a giant strength similar to that in typical optoelectronic materials like GaAs  \cite{Bechstedt.Belabbes-2013-JoPCM}.

\begin{table}[h!]
\begin{ruledtabular}
\caption{Normalized orbital character of the lowest conduction band states $\Gamma_{8c}$ characterized by the orbital- and
site-projected character of the wave function for hex-Si$_{0.25}$Ge$_{0.75}$, 2H-Ge, and structurally perturbed hex-Ge. In the case of hex-Si$_{0.25}$Ge$_{0.75}$
the fourth atom is Si.
}
\centering
\begin{tabular}{ccccc} 
              & \multicolumn{3}{c}{orbital}     \\ \cline{2-4}  
atom    & $s$                       & $p_z$                    & $d_{z^2}$ \\  \hline
1           & 0.92, 0.70, 0.71 & 0.02, 0.20, 0.13 & 0.06, 0.10, 0.16 \\
2           & 0.83, 0.70, 0.65 & 0.14, 0.20, 0.24 & 0.03, 0.10, 0.10 \\
3           & 0.51, 0.70, 0.75 & 0.24, 0.20, 0.13 & 0.26, 0.10, 0.12 \\
4           & 0.41, 0.70, 0.70 & 0.48, 0.20, 0.21 & 0.12, 0.10, 0.09 \\
average & 0.75, 0.70, 0.71 & 0.17, 0.20, 0.18 & 0.09, 0.10, 0.11 
\label{tab2}
\end{tabular}
\end{ruledtabular}
\end{table}

It is possible that structural and chemical perturbations may explain why in nanostructured Ge core-shell wires strong emission is experimentally observed already for nominally pure hexagonal Ge. Similar theoretical and experimental observations have been made for pseudodirect wurtzite semiconductors in comparison
to indirect zinc-blende materials, e.g. for InP nanowires alloyed with Al \cite{Gagliano.Kruijsse.ea:2018:NL}, or in wurtzite GaP nanowires, where strong many-body excitonic effects
or tensile uniaxial strain  may explain the measured strong luminescence \cite{Belabbes.Bechstedt:2019:PSSB,Greil.Assali.ea:2016:NL,Gagliano.Belabbes.ea:2016:NL}.

For a better understanding of the origin of the increased oscillator strengths we have separately considered the effect of a purely structural perturbation versus combined chemical and structural perturbation on lonsdaleite Ge.
Considering the results in Fig.~\ref{fig3} and Table~\ref{tab2}, we can conclude that the influence of the structural perturbation is much weaker than the chemical effect,
despite the fact that Si and Ge are isovalent. The structural relaxation increases the interband transition
energies in Fig.~\ref{fig1}(b) and the oscillator strengths in Fig.~\ref{fig3}. However, the effects are smaller as in the case of a concurrent chemical and structural modification, as in the hexagonal alloy Si$_{0.25}$Ge$_{0.75}$. The small changes of the momentum matrix elements are thereby in line with the minor variation
of the orbital character in Table~\ref{tab2}. The small effect of purely structural modifications cannot be generalized. In the case
of moderate tensile uniaxial strain, as shown for 2H-Ge in Fig.~\ref{fig4}, we can still report weak
or moderate changes of the dipole strengths of the transitions $\Gamma_{9v}$, $\Gamma_{7+v}$, $\Gamma_{7-v}\rightarrow\Gamma_{7c}$,
$\Gamma_{8c}$. In fact, in contrast to the band energies, the wave functions experience minor modifications under strain, and therefore the absolute values of the oscillator strenghts are nearly preserved. Following the trend with rising tensile uniaxial strain \cite{Suckert.Roedl.ea:2021:PRM}, an inversion of the two lowest conduction bands occurs, i.e., the $\Gamma_{7c}$ conduction band appears below the $\Gamma_{8c}$
band at $\Gamma$. This band inversion leads to huge oscillator strengths for optical transitions to or from the lowest conduction band, similarly to what is shown in Fig.~\ref{fig3}
for 2H-Ge and the second conduction band. A similar band inversion $\Gamma_{7c}\Leftrightarrow\Gamma_{8c}$ is 
suggested to happen in the band structure of the 4H polytype \cite{Raffy.Furthmueller.ea:2002:PRB} of Ge  
\cite{Kiefer.Hlukhyy.ea:2010:JMC}.
Our own test calculations for 4H-Ge do not give rise to such a band inversion nor, consequently,
strong optical transitions to or from the lowest conduction band. This finding is in agreement with the fact that the pure hexagonal
stacking in 2H is mixed with a cubic stacking in the 4H polytype \cite{Raffy.Furthmueller.ea:2002:PRB}.

\section{Conclusions} 

In summary,  we have clearly demonstrated by means of first-principles calculations that lonsdaleite germanium underlies a transition
from a pseudo-direct semiconductor with a dipole-allowed but very weak optical oscillator strength into a direct semiconductor with
typical strength of a $p\rightarrow s$ transition, upon perturbation of the ideal hexagonal lattice. Here we consider the substitution
of one Ge atom with a Si atom in the four atom unit cell. The alloying of 2H-Ge with Si leads not only 
to an increase of the interband transition energies but also to the enhancement of the dipole-matrix elements of all optical transitions from the three highest
valence bands into the lowest conduction band, independently of light polarization. The increase of the oscillator strengths at $\Gamma$ leads to matrix elements of the order of those for
transitions involving the second conduction band. Strong oscillator strengths of the lowest interband transitions also occur in uniaxially strained 2H-Ge because of the order inversion of the two lowest conduction bands. All in all, we conclude that perturbed lonsdaleite Ge is an excellent material for active optoelectronic applications in light-emitting diodes and lasers.
\vspace{-0.3 cm} 

\section{\label{sec2} Computational Methods}

The atomic configurations and lattice constants of the hexagonal crystals are optimized applying density functional theory (DFT) as implemented in the Vienna Ab-initio Simulation Package (VASP) \cite{Kresse.Furthmueller:1996:PRB} using the
projector-augmented wave method \cite{Kresse.Joubert:1999:PRB} and a plane-wave cutoff of 500~eV. The shallow Ge3$d$ electrons are treated as valence electrons. The modified Perdew-Becke-Ernzerhof (PBEsol) XC functional \cite{Perdew.Ruzsinszky.ea:2008:PRL} is applied for structural relaxations. The BZ integration is performed using a $\Gamma$-centered
12$\times$12$\times$6 {\bf k}-point grid. Atomic geometries are relaxed until the Hellmann-Feynman forces are below
1~meV/{\AA}. The spin-orbit interaction is included for all electronic structure calculations.
\vspace{-0.3 cm} 
\section*{Acknowledgments}

We acknowledge financial support from the H2020-FETOpen projects SiLAS (grant agreement No. 735008) and OptoSilicon (grant agreement No. 964191).
\vspace{-0.3 cm} 
\section*{Conflict of Interest}
The authors declare no conflict of interest.
\vspace{0.3 cm} 
\section*{Data Availability Statement}
The data that support the findings of this study are available on reasonable request from the corresponding author.
\vspace{0.1 cm} 
\section*{Keywords}
germanium, silicon-germanium alloys, lonsdaleite, electronic structure, optical matrix elements, light emission


\bibliography{literatur1}
\end{document}